\documentclass[aps,prb,citeautoscript,twocolumn,superscriptaddress,footinbib,10pt,longbibliography,nobibnotes]{revtex4-1} 
\usepackage[utf8x]{inputenc}
\usepackage{amsmath,amssymb}
\usepackage{mathrsfs} 
\usepackage{graphicx,color,colortbl}
\usepackage{rotating,array,tabularx,booktabs}
\usepackage{varwidth,xcolor}
\usepackage{placeins}
\usepackage{siunitx}
\usepackage[normalem]{ulem}

\DeclareGraphicsExtensions{.pdf,.PDF,.png,.PNG}

\makeatletter
\newcommand*{\balancecolsandclearpage}{%
  \close@column@grid
  \clearpage
  \twocolumngrid
}
\makeatother

\allowdisplaybreaks

\newcommand{\dif}{\mathrm{d}}%
\newcommand{\Nabla}{\vec{\nabla}}%
\newcommand{\abs}[1]{\lvert#1\rvert}%
\newcommand{\norm}[1]{\lVert#1\rVert}%
\newcommand{\ZT}[1]{\textquotedblleft#1\textquotedblright}%

\renewcommand{\vec}[1]{\boldsymbol{#1}}%
\newcommand{\eqdot}{\,.}
\newcommand{\eqcomma}{\,,}

\begin{document}

\title{A colloidal time crystal and its tempomechanical properties}
\author{Marina Evers}
\affiliation{Institut f\"ur Theoretische Physik, Center for Soft Nanoscience, Westf\"alische Wilhelms-Universit\"at M\"unster, D-48149 M\"unster, Germany}

\author{Raphael Wittkowski}
\email[Corresponding author: ]{raphael.wittkowski@uni-muenster.de}
\affiliation{Institut f\"ur Theoretische Physik, Center for Soft Nanoscience, Westf\"alische Wilhelms-Universit\"at M\"unster, D-48149 M\"unster, Germany}

\begin{abstract}
The spontaneous breaking of symmetries is a widespread phenomenon in physics. When time translational symmetry is spontaneously broken, an exotic nonequilibrium state of matter in which the same structures repeat themselves in time can arise. This state, known as \ZT{time crystal}, attracted a lot of interest recently. Another relatively new research area deals with active matter. Materials consisting of colloidal particles that consume energy from their environment and propel themselves forward can exhibit intriguing properties like superfluidity that were previously known only from quantum-mechanical systems. Here, we bring together these -- at first glance completely different -- research fields by showing that self-propelled colloidal particles can form classical continuous time crystals. We present a state diagram showing where this new state of matter arises. Furthermore, we investigate its tempomechanical properties and present a temporal stress-strain diagram showing parallels to conventional materials science but also remarkable new properties that do not have a counterpart in common materials.
\end{abstract}

\maketitle

\section{Introduction}
The proposal of time crystals\cite{ShapereW2012, wilczek2012quantum} as a theoretical concept in 2012 opened a new area of research \cite{ZhangEtAl2017, ChoiEtAl2017, sacha2017time,khemani2019brief, khemani2017defining, iemini2018boundary, bruno2013impossibility, WatanabeO2015, autti2020AC, else2016floquet}. A time crystal is characterized by spontaneously breaking time-translational symmetry and forms an analog to spatial crystals that break translational symmetry in space. Similar to spatial crystals, time crystals form a state of matter. 
Soon, it was proven that time crystals cannot exist in equilibrium \cite{bruno2013impossibility, WatanabeO2015}. Thus, time crystals are a non-equilibrium phenomenon and never reach thermal equilibrium. Around the same time, the concept of discrete time crystals, which describe a discrete breaking of time translational symmetry in periodically driven systems, was proposed. While continuous time crystals break continuous time-translational symmetry, discrete time crystals occur in periodically driven systems but oscillate at an integer fraction of the driving frequency. Thus, they show a discrete time-translational symmetry breaking. Their existence in quantum mechanical systems was experimentally proven for the first time in 2017 \cite{ZhangEtAl2017, ChoiEtAl2017}. Recently, different suggestions for (discrete) classical time crystals were discussed \cite{yao2020classical,das2018cosmological,heugel2019classical,gambetta2019classical,kessler2021observation}. However, the existence of classical continuous time crystals is still under debate.

Active colloidal particles are nano- or microparticles characterized by the capability to consume energy from their environment and to use it for self-propulsion \cite{BechingerdLLRVV2016}. These particles include a large variety of motile microorganisms, such as archaea, bacteria, and many eukaryota. On the other hand, many types of artificial microswimmers, that utilize, e.g., acoustophoresis, chemophoresis, thermophoresis, or shape deformations for self-propulsion, have been developed in recent years \cite{BechingerdLLRVV2016, marchetti2013hydrodynamics, romanczuk2012active, wang2013small,duan2015synthetic,BechingerdLLRVV2016}. 
Such active particles have been found to exhibit a fascinating nonequilibrium behavior where various exotic properties can arise. Some of these properties, such as superfluidity, were previously known only from quantum systems and other properties, such as a negative viscosity, were previously unknown from both classical passive (i.e., not active) systems and quantum systems \cite{saintillan2018rheology, lopez2015turning}. 
Active colloidal particles have also been shown to give rise to a complex state diagram with various unusual nonequilibrium states including active crystals \cite{digregorio2018full, marchetti2016minimal}, which are colloidal crystals formed by active particles. Lately, there have been initial studies investigating the formation and stability of active colloidal crystals \cite{bialke2012crystallization, briand2016crystallization, menzel2014active, klamser2019kinetic, digregorio20192d, digregorio2019clustering, briand2018spontaneously}. However, the properties of this state of active matter remain mostly unexplored up to now. 

Recently, the interaction of colloids with a choreographic time-crystalline lattice formed by traps was studied\cite{libal2020colloidal}. Also, a first connection between an active fluid and a time crystal was proposed \cite{fruchart2020phase}. However, this proposal was based on an abstract field-theoretical model, that can be related to active fluids but is not applicable to active crystals, and the study did not demonstrate the emergence of a time-crystalline state for a specific and realistic system of active particles. 

In the following, we establish a concrete connection between active colloidal crystals and time crystals. We show for a specific and realistic system that active colloidal crystals can form classical continuous (and potentially even living) time crystals. Furthermore, we study the state diagram of these time crystals and investigate their tempomechanical properties based on a temporal stress-strain diagram.

\section{\label{sec:Results}Results}
\subsection{\label{sec:periodicMotion}Periodic motion in active crystallite}
To obtain new insights into the properties of active colloidal crystals, we use computer simulations combining low-Reynolds-number hydrodynamics with Brownian particle dynamics (see Methods for details). 
 
We investigate a small active crystallite consisting of seven spherical particles with radius $R$ that are placed in form of a hexagon with one central particle and surrounded by a liquid. The distance $d$ between the particles is set to $2.2R$ and they are fixed in their positions but can freely change their orientations due to hydrodynamic interactions. 
Each swimmer creates a flow field caused by its swimming motion. This flow field influences the other swimmers and causes them to rotate.
 
The particles themselves are modelled as squirmers, which today is a standard model for active swimmers\cite{lighthill1952squirming}.
Depending on the flow field generated by the squirmers, they are distinguished into pullers, which propel themselves mainly with the front similar to the algae \textit{Chlamydomonas reinhardtii}, neutral squirmers, where the swimming motion happens everywhere on the body like for \textit{Paramecia}, or as pushers, which propel themselves with the back part of their body like \textit{Escherichia coli} \cite{BechingerdLLRVV2016}. 

\begin{figure*}[htb]
\centering\includegraphics[width=\linewidth]{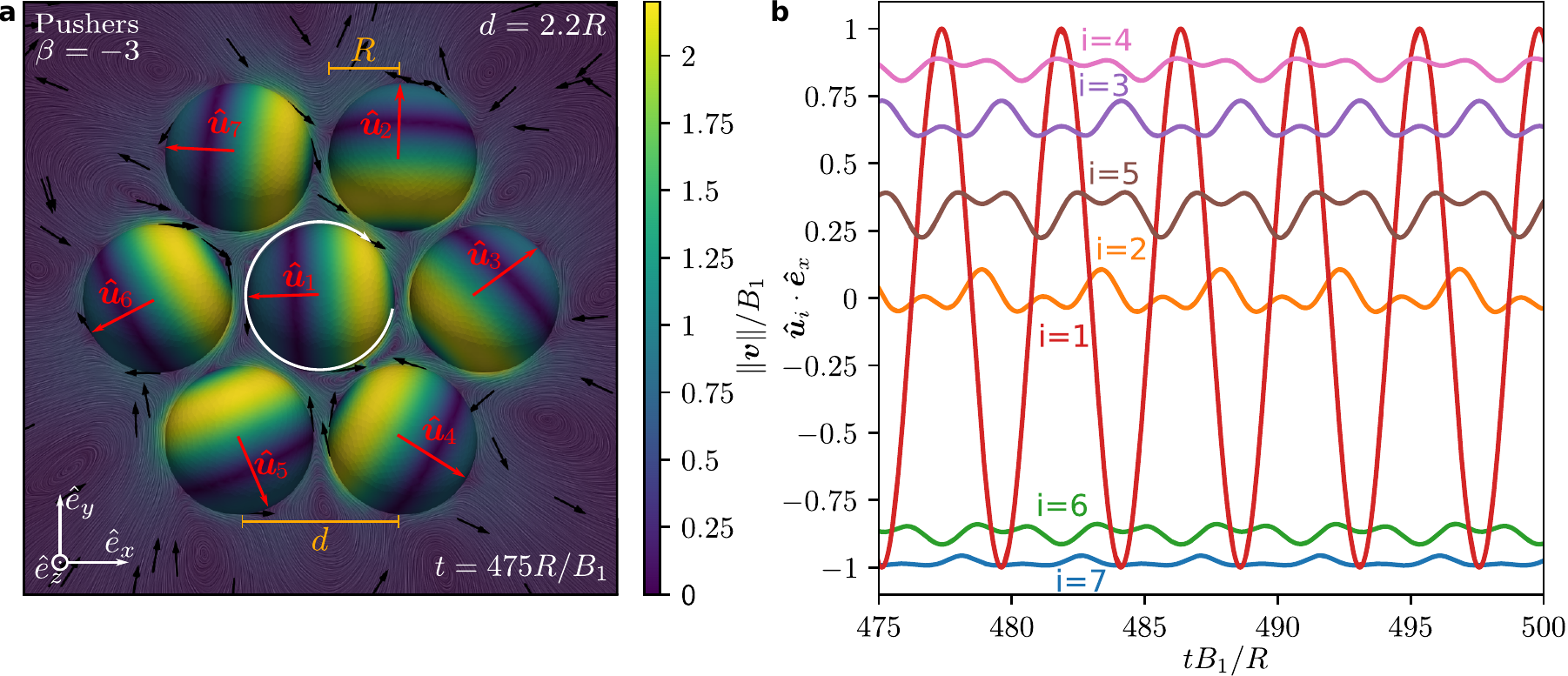}
\caption{\label{fig:1}\textbf{Self-organized time-periodic motion of active colloidal particles.} \textbf{a} If the particles are pushers with radius $R$, distance $d=2.2R$, and squirming parameter $\beta=-3$, they self-organize an orientational configuration that behaves periodically in time and constitutes a \textit{colloidal time crystal}. The snapshot shows the particle orientations $\vec{\hat{u}}_i$ with $i=1,\dotsc,7$ and the flow field $\vec{v}$ around the particles at time $t=475R/B_1$, where $B_1$ denotes the first velocity mode. 
\textbf{b} As the time evolution of the particle orientations shows, the central particle ($i=1$) rotates clockwise with a constant speed, whereas the other particles oscillate periodically.}
\end{figure*}

For an active crystallite formed by pushers with squirming parameter $\beta=-3$, we observed a periodic rotational motion after an initial transition phase (see Fig.\ \ref{fig:1}). 
It is mainly characterized by the rotation of the central particle which rotates clockwise. 
The other six particles only slightly change their orientation, but this motion follows the same periodicity as the one of the central particle. 
In this state of periodic rotational motion, all particles lie in the $x$-$y$-plane and the particle directors, which denote the orientations of the particles, do not leave this plane for the chosen parameter setting. 

The plane forms a fixed point of the system, as can be seen by symmetry considerations. Thus, in absence of noise the particles should not rotate out of the plane once all particles reached it. We do not impose planar swimming directions as initial conditions, but the system self-organizes into the planar motion. This suggests that this fixed point is stable. We investigate its stability in more detail in Section~\ref{sec:interactions}.

\begin{figure*}[htb]
\centering\includegraphics[width=\linewidth]{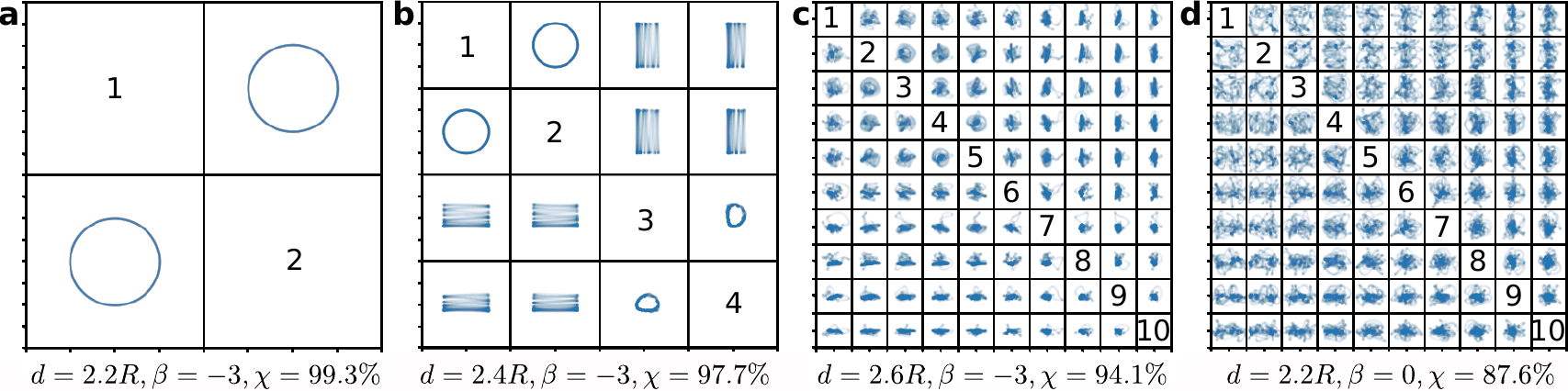}
\caption{\label{fig:pca}\textbf{PCA projections of the 21-dimensional trajectory $\boldsymbol{\vec{u}(t)}$ describing the whole system.} 
The axes of each subplot range from $-2$ to $2$. $\chi$ denotes the percentage of variation covered by the chosen number of eigenvalues. \textbf{a} The time-crystalline state is described by a two-dimensional circular trajectory. \textbf{b} When $d$ increases, the trajectory leaves the two-dimensional plane. \textbf{c} For even larger values of $d$, no order is visible and more components occur. \textbf{d} The trajectory describing a crystallite of neutral squirmers ($\beta=0$) also shows no order.}
\end{figure*}

For a more general analysis including all degrees of freedom of the particles, we combine their seven orientation vectors (i.e., directors) $\vec{\hat{u}}_i(t)$ with $i=1,\dotsc,7$ and time $t$ to a 21-dimensional vector $\vec{u}(t)=(\vec{\hat{u}}_1(t),\dotsc,\vec{\hat{u}}_7(t))^\mathrm{T}$ that describes the state of the whole system at a certain time $t$. We project the 21-dimensional trajectory to a lower-dimensional space using a principal component analysis (PCA)\cite{jolliffe2016principal}. For this periodic state, we found that the first two principal components already contain $\chi = 99.3\%$ of the variation in the data set. By visualizing these first two components (see Fig.~\ref{fig:pca}), we obtained a spherical motion. Thus, the system can be described by a simple two-dimensional spherical motion in the 21-dimensional phase space. 
 
This periodic motion spontaneously breaks the continuous time-translational symmetry in a way that is similar to the breaking of translational symmetry occurring in an ordinary crystal in space. Analogous to the periodicity in space that is a characteristic of a normal crystal, our system is periodic in time (see Fig.\ \ref{fig:1}), which is typical for a time crystal. This is especially interesting as the active particles are no chiral particles but the self-propulsion of a free particle has an axis of rotational symmetry that points through the center of the spherical particle.

To call a system a time crystal and distinguish it from other well-known nonequilibrium phenomena that show periodicity, like oscillating chemical reactions or convection cells, it has to fulfill four criteria in addition to the time-translational symmetry breaking\cite{yao2020classical}, namely independence on fine-tuning of parameters, stability against fluctuations, periodicity arises from interactions, and periodicity holds in larger systems.
As we show now, all of these criteria are fulfilled by the active crystallite studied in the present article. 

\subsection{\label{sec:stateDiagram}State diagram}
First, time crystallinity forms a state of matter and, thus, should not require fine-tuning the system's initial conditions or parameters. 
We confirmed that this is the case here by performing the simulations for different, randomly set initial conditions and various parameter combinations (see Methods for details). 

\begin{figure}[htb]
\centering\includegraphics[width=\linewidth]{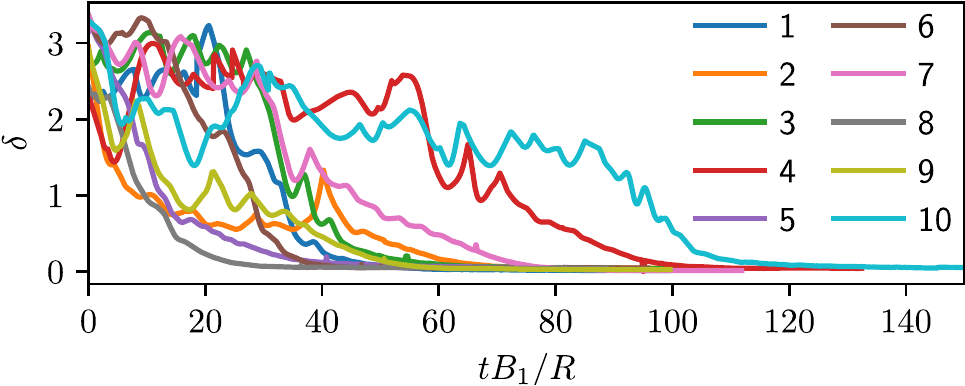}
\caption{\label{fig:initialConditions}\textbf{Influence of initial conditions.} The distance $\delta(t)$, which measures how close the state of the system at later times comes back to its state at time $t$, is shown as a function of $t$ for $d=2.2R$, $\beta=-3$, and $10$ different initial conditions. In all cases, the behavior converges towards a periodic state.}
\end{figure}

We start with the initial conditions. Repeating the simulations that showed the periodic state described in Section \ref{sec:periodicMotion}, we observed that the length of the transition phase varied, but in all simulations the swimmers showed the same qualitative steady-state behavior as depicted in Fig.\ \ref{fig:1}. 
To verify this finding more quantitatively, we introduce a distance measure
\begin{equation}
\begin{split}%
\delta(t) &= \min\!\Big(\min_{0<r<m_1}\norm{\vec{u}(t) - \vec{u}(r)} \eqcomma\\ 
&\quad\,  \min_{m_2<r<t_{\max}}\norm{\vec{u}(t) - \vec{u}(r)}\Big) \eqcomma
\end{split}%
\label{eq:distance}%
\end{equation}
where $m_1$ is the closest local maximum of the $r$-dependent function $\norm{\vec{u}(t) - \vec{u}(r)}$ to $t$ with $m_1 < t$, $m_2$ is the closest local maximum with $m_2 > t$, and $t_{\max}$ is the last simulated point in time. 
Thus, $\delta(t)$ describes how similar the system's state at time $t$ is to its state at another time. For periodic motion, the trajectory should come back to a given point for every time and, therefore, $\delta(t)=0$ should apply for different values of $t$. For non-periodic cases, however, $\delta(t)$ is larger than zero, as the trajectory never reaches any original point again. With this definition, we can define the end of the initial transition phase for periodic motions as the point where $\delta(t)$ becomes zero. In Fig.~\ref{fig:initialConditions} we plot the distance measure $\delta(t)$ for $10$ different initial conditions.
We see that after some transition phase, the distance decreases to zero for the different initial conditions. This confirms that the occurrence of a periodic state does not rely on particular initial conditions.

\begin{figure*}[htb]
\centering\includegraphics[width=\linewidth]{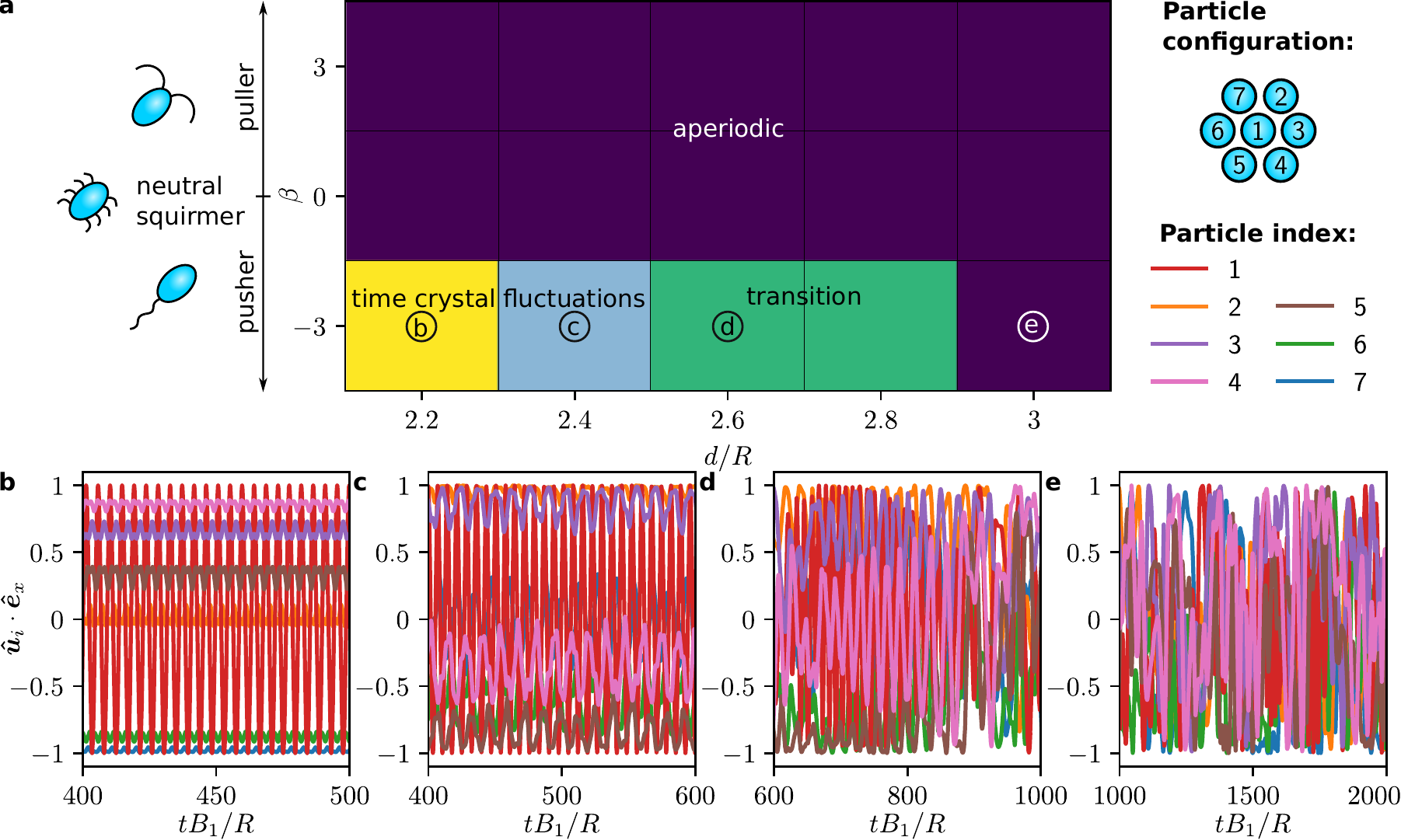}
\caption{\label{fig:2}\textbf{State diagram of an active-particle system showing a time-crystalline region.} \textbf{a} The state diagram of the system from Fig.\ \ref{fig:1} for variable particle distance $d$ and squirming parameter $\beta$ involves four different states including a time crystal. \textbf{b-e} For each state, a representative part of the time evolution of the particle orientations is shown. The supplemental videos V1-V4 show the time evolution of the flow field for each of the four states.}
\end{figure*}

To analyze how the state of the system depends on the values of its parameters, we now investigate the system's state diagram (see Fig.~\ref{fig:2}), which shows the behavior of the active crystallite for particle distances $d=2.2R,2.4R,\dotsc,3R$ and squirming-parameter values $\beta=-3,0,3$ that correspond to different types of swimmers (pushers for $\beta<0$, neutral squirmers for $\beta=0$, pullers for $\beta>0$).

 \begin{figure*}[htb]
\centering\includegraphics[width=\linewidth]{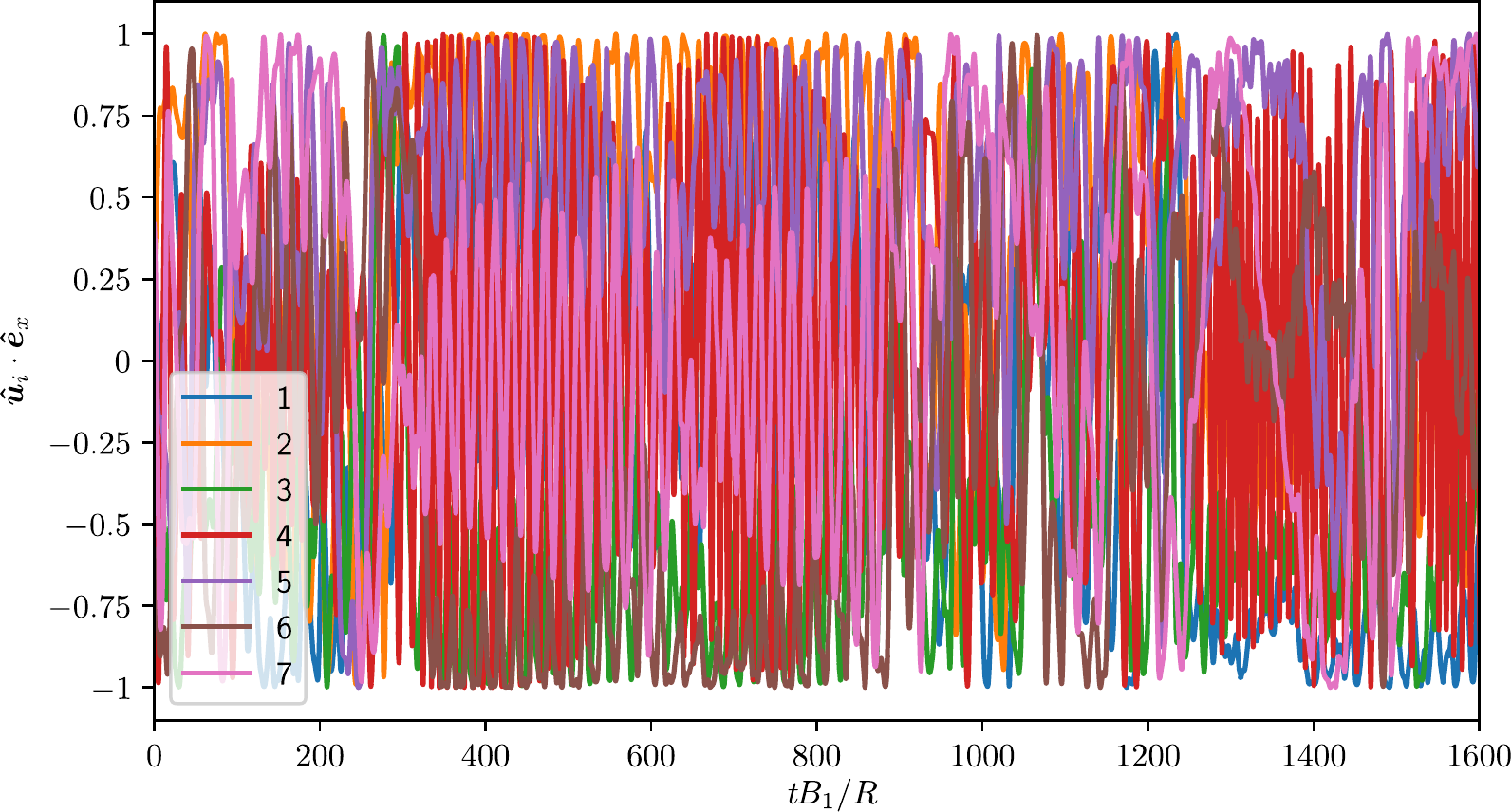}
\caption{\label{fig:3}\textbf{Extended diagram for the transition state shown in Fig.\ \ref{fig:2}.} This diagram corresponds to an active crystallite formed by pushers with distance $d=2.6R$ and squirming parameter $\beta=-3$.}
\end{figure*}

Interestingly, self-organization and time crystallinity are observed only for pushers, whereas for pullers and neutral squirmers only chaotic-looking aperiodic motion is found (see Fig.\ \ref{fig:2}a). This shows that the type of interaction between the particles is crucial for forming a time crystal (see Section~\ref{sec:interactions} for details).

In the case of pushers, different states occurring for different particle distances can be distinguished. 
Time crystallinity with periodic motion of the particles is observed for the smallest considered particle distance $d=2.2R$ (see Fig.\ \ref{fig:2}b). 
For a larger particle distance of $d=2.4R$, the motion is almost periodic but with clearly visible fluctuations (see Fig.\ \ref{fig:2}c).  
When the particles have an even larger distance $d=2.6R$ or $d=2.8R$, the motion looks more irregular and suggests a transition. 
At some times, it seems as if the system is converging towards a periodic state, but never reaches it completely; 
at other times, the motion looks chaotic (see Fig.\ \ref{fig:2}d and, for a larger time frame, Fig.\ \ref{fig:3}). When comparing the latter state to classical systems, we can interpret it as a dynamical coexistence between ordered and unordered (temporal) regions. This reveals parallels to a coexistence of a liquid phase and a crystalline phase. We therefore consider this state as a coexistence state.
For the largest considered particle distance $d=3R$, aperiodic motion similar to the behavior of pullers and neutral squirmers is observed (see Fig.\ \ref{fig:2}e).
The transition from a time crystal to an aperiodic motion for increasing particle distance can be explained by the decrease in
the hydrodynamic interaction between the particles for increasing particle distance. 

\begin{figure*}[htb]
\centering\includegraphics[width=\linewidth]{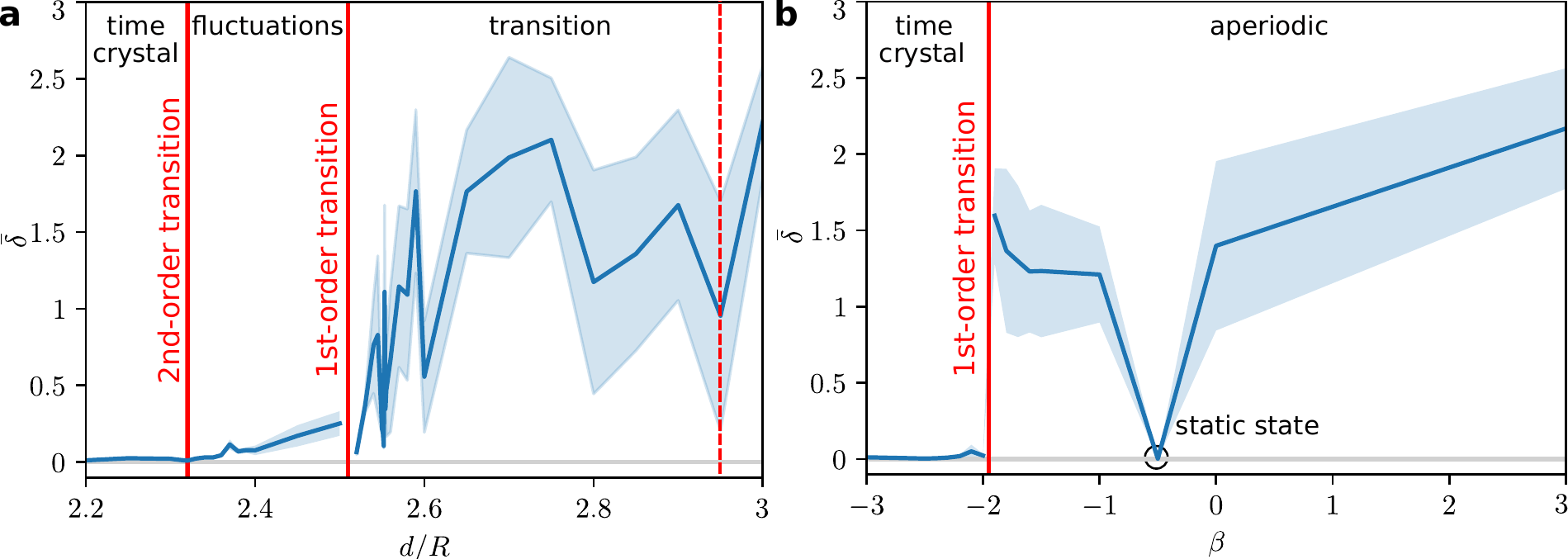}
\caption{\label{fig:phaseTransition}\textbf{Transitions between the states from Fig.\ \ref{fig:1}a.} The solid blue curve represents the value of the order parameter $\bar{\delta}$, whereas the shaded region shows the standard deviation of $\delta(t)$. \textbf{a} Variation of the particle distance $d$ for $\beta=-3$. There is a second-order transition at $d=2.32R$ and a first-order transition at $d=2.51R$. At $d=2.95R$, we found by visual inspection a change from the transition state to the fully aperiodic state that is not visible in the order parameter $\bar{\delta}$. \textbf{b} Variation of the squirming parameter $\beta$ for $d=2.2R$. The system shows a first-order transition at $\beta=-2$. Note that at $\beta=-0.5$ the order parameter vanishes due to a static state of the system.}
\end{figure*}

To better understand how the system's full trajectory $\vec{u}(t)$ varies between the different states, we observe the PCA results for the individual states as shown in Fig.~\ref{fig:pca}. The circular structure in phase space of the time-crystalline state (see Fig.\ \ref{fig:pca}a) disappears when $d$ or $\beta$ is increased. We can see that the dimensionality of the curve in phase space increases and the ordered motion in the first principal component vanishes. While we can represent $\chi = 97.7\%$ of the motion in four principal components for $d=2.4R$, we already need ten components for $\chi = 94.1\%$ of the variation in case of $d=2.6R$. In the latter case, even the first component shows no order. For $\beta=0$, the first ten components only cover $\chi = 87.6\%$ of the variation without any visible order.
 
For studying the transitions between the observed states, we define an order parameter $\bar{\delta}$ as the time average of the minimum distance of the 21-dimensional system trajectory $\vec{u}(t)$ to itself:
\begin{equation}
\bar{\delta} = \frac{1}{t_{\max}}\int_0^{t_{\max}} \delta(t)\, \dif t\eqdot
\end{equation}
Here, $\delta(t)$ is the distance measure introduced in Eq.\ \eqref{eq:distance}. 
Thus, $\bar{\delta}=0$ applies to periodic states and $\bar{\delta}>0$ to non-periodic states. 
Compared to frequency-based order parameters, this one allows to capture gradually changing behavior, if the system's trajectory is not periodic but moves in a confined region of the parameter space. For the calculation of the transitions, we run simulations starting with the same initial conditions but with varied values for $d$ and $\beta$. We refine the sampling of the parameter range in regions of a transition. All simulations are run over a time span of $1000R/B_1$. For the calculation of the order parameter, we use the results of the last $500R/B_1$ to exclude the transition phase.

The simulation results for a variation of the particle distance $d$ are shown in Fig.~\ref{fig:phaseTransition}a. Until a distance of $2.32R$, we see a periodic state which corresponds to an order parameter of $\bar{\delta}=0$. It is followed by a continuous, second-order transition with a critical exponent of $0.35$. For an increasing distance, up to $d=2.51R$, we can see fluctuations in the particles' trajectories which result in a growing but still relatively small order parameter. At $d=2.51R$, a jump in the order parameter occurs which can be characterized as a discontinuous, first-order transition. For larger distances, there is no visible order in the particles' trajectories which results in fluctuating, larger values of $\bar{\delta}$. However, when observing the individual trajectories of the system, we can distinguish between the transition state and the fully aperiodic state as shown in Figs.\ \ref{fig:2}d and \ref{fig:2}e, respectively. The boundary between both states can be located at about $d=2.95R$. However, this transition between both states is not visible in the curve for the order parameter, which is consistent with our previous finding that the former of the two states resembles rather a coexistence state than an individual phase.

The evolution of the order parameter $\bar{\delta}$, when $\beta$ is varied, is depicted in Fig.~\ref{fig:phaseTransition}b. Here, we can identify a periodic behavior for strongly negative values of $\beta$ and a discontinuity at $\beta=-2$. This discontinuity can be identified as a first-order transition. For $\beta=-0.5$, we reach a special case. There, the system converges towards a static state which also manifests itself in an order parameter of $\bar{\delta}=0$. However, by observing the orientations of the individual particles, we can confirm that no periodic motion but a static state occurs there. In this state, all particles but the central one are oriented in the same direction. The central particle aligns with the $z$-axis of the system.

To further study the transitions, we considered calculating hysteresis curves as it is common in classical (spatial) systems. However, this involves the difficulty that both the change of a driving parameter for the hysteresis curve as well as the observed behavior are temporal variations. Thus, each hysteresis curve would strongly depend on the speed of change in the driving parameter and yield different and possibly misleading results. This reveals that not all methods used to study spatial crystals can be directly transferred to the study of time crystals.
 
The previous results show that fine-tuning either the initial conditions or the systems parameters is not required and distinguishes our system from other well known nonlinear systems that show periodic behavior.

\subsection{Influence of fluctuations}
Second, the system should be stable against fluctuations. 
All simulations are influenced by numerical errors that are caused by the chosen granularity of the underlying mesh and by rounding errors. As is it usual, we ensured that the limited resolution has no significant influence on our simulation results by performing comparative simulations with a finer discretization. Even though the numerical errors are small, they can influence the system's behavior at large time scales. Although the system's symmetry should prevent particles from moving out of the particle plane, they do this in some cases for parameter combinations that correspond to non-periodic states. For the periodic state, however, we did not observe such a behavior, suggesting that this state is stable against numerical fluctuations.

\begin{figure*}[htb]
\centering\includegraphics[width=\textwidth]{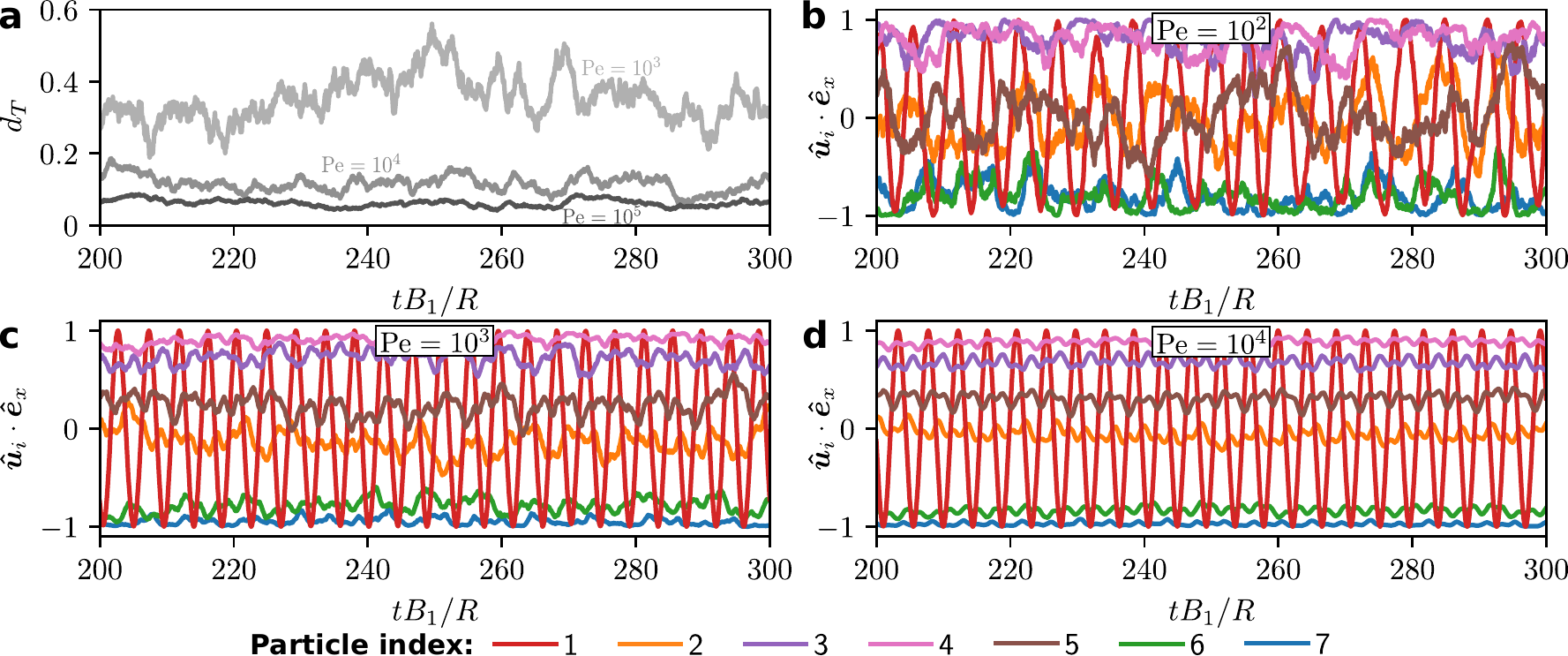}
\caption{\label{fig:addedBrownian}\textbf{Influence of Brownian motion.} All plots correspond to the observed time-crystalline state with parameter values $d=2.2R$ and $\beta=-3$. \textbf{a} The Euclidean distance $d_T$ between the undisturbed trajectories and the trajectories with Brownian motion decreases for an increasing P\'eclet number. To eliminate shifts in time, dynamic time warping \cite{salvador2007toward} was applied before calculating the distance. \textbf{b} The trajectories of single particles for $\mathrm{Pe}=100$ (not included in (\textbf{a})) are strongly disturbed but the central particle (particle 1) shows a similar motion as in the undisturbed case. \textbf{c, d} For increasing $\mathrm{Pe}$, the influence of the Brownian motion becomes smaller and the motion of the particles approximates the undisturbed motion.}
\end{figure*}

Heating of the system is prevented by coupling to a bath introducing friction and Brownian noise. We therefore studied also the influence of Brownian motion on the system's behavior. For this purpose, we included Brownian motion in the particles' equations of motion. Brownian motion is commonly introduced in active systems consisting of colloidal particles, since their motion is influenced by collisions with solvent molecules. 
We study the influence of Brownian motion as a function of the dimensionless P\'eclet number $\mathrm{Pe}=B_1/(RD_R)$, where $B_1$ denotes the surface velocity mode of the active particles that also determines the swimming velocity and $D_R$ is their rotational diffusion coefficient. The results for different values of $\mathrm{Pe}$ are shown in Fig.~\ref{fig:addedBrownian}. We can see that the overall motion of the system is robust against Brownian motion if it is small compared to the propulsion of the particles. The Euclidean distance $d_T$ between the system's trajectory with Brownian motion and the corresponding trajectory without Brownian motion is presented in Fig.~\ref{fig:addedBrownian}a. To take shifts in time into account, we apply dynamic time warping \cite{salvador2007toward} before calculating the distances. 
Overall, the trajectories get closer to the undisturbed trajectory if $\mathrm{Pe}$ increases. This follows the expectation as an increasing value of $\mathrm{Pe}$ means a lower influence of the Brownian motion. Even the motion of particles with $\mathrm{Pe}=100$ shows similarities to the motion of the undisturbed system (see Fig.~\ref{fig:addedBrownian}b), albeit the influence of the Brownian motion is clearly visible. Furthermore, we can see that the trajectories for $\mathrm{Pe}\geq1000$ only show small deviations. Comparing the motion of the individual particles confirms that the characteristic behavior is preserved with Brownian motion (cf.\ Fig.~\ref{fig:addedBrownian}c,d). Especially the rotating motion of the central particle is not influenced, whereas the influence of the fluctuations on other particles is larger.

\subsection{\label{sec:interactions}Particle-particle interactions}
Third, the time periodicity should only arise from the interaction between the particles, i.e., the degrees of freedom should be coupled locally. This also applies to our system, where the hydrodynamic interactions are crucial. Without interactions between the particles, they would not move at all.

We study the interactions between two particles in the system and their dependence on the orientation angle $\alpha$ of a particle, which is measured counterclockwise within the particle plane (i.e., the $x$-$y$-plane) relative to the position of the other considered particle, as well as on the parameters $d$ and $\beta$ to better understand how the dynamic states emerge and why transitions between the states occur. This also helps to understand how the propulsion type influences the system's behavior. 
We observe the influence of a particle of the system on itself and on the other particles. As the time-crystalline behavior is a planar motion and the motion in the plane forms a fixed point in the dynamics of the system, we consider both the torque $T_z$ exerted on a particle that leads to a rotation within the particle plane and the torque $T_\mathrm{rot}$ that rotates the particle out of or back into the plane.
\begin{figure*}[htb]
\centering\includegraphics[width=\linewidth]{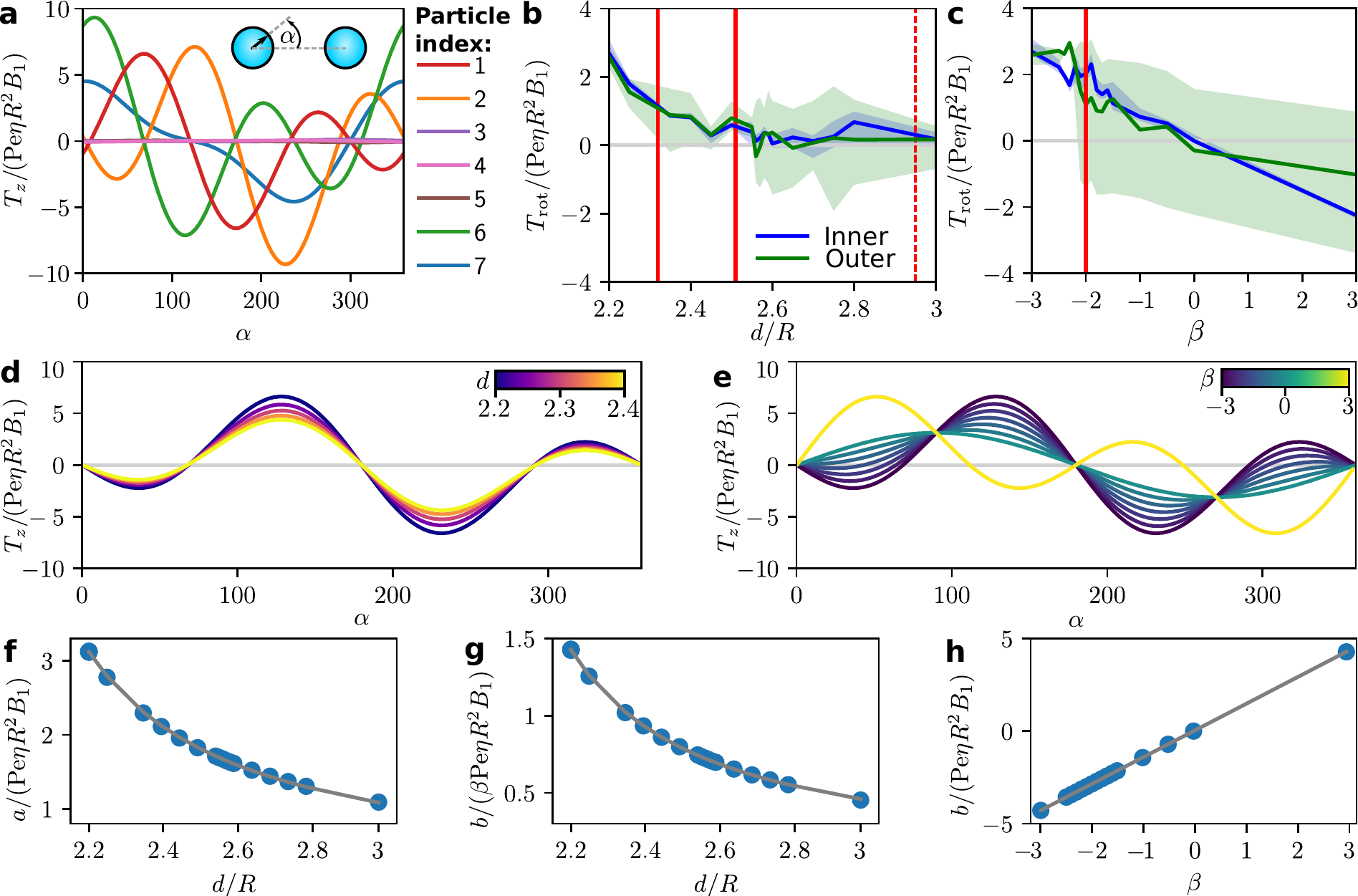}
\caption{\label{fig:addedInteraction}\textbf{Analysis of interactions.} Unless stated otherwise, the parameters are $d=2.2R$ and $\beta=-3$. \textbf{a} Torque $T_z$ exerted by the propulsion of an outer particle (particle 7) on the other particles and itself. The nearest neighbors (1, 2, 6) are strongly influenced whereas the second nearest neighbors (3, 4, 5) only experience very small torques. \textbf{b, c} Torque $T_\mathrm{rot}$ that causes the inner particle (particle 1) or an outer particle, when its orientation deviates by \SI{2}{\degree} from the plane, to rotate itself back to the plane (positive) or further away from the plane (negative) as a function of the parameters $d$ and $\beta$. The curves represent the torque averaged over all in-plane orientation angles $\alpha$ and the shaded areas show the corresponding range of values. Note that these curves only include the values of $\alpha$ that occurred in the simulations for Fig.~\ref{fig:phaseTransition}. The vertical lines indicate boundaries of states and correspond to those from Fig.\ \ref{fig:phaseTransition}. \textbf{d, e} Torque $T_z$ exerted by the propulsion of the inner particle on one of the other particles as a function of the angle $\alpha$ and for different values of the parameters $d$ and $\beta$. \textbf{f-h} Dependence of the fit parameters $a$ and $b$ from Eq.\ \eqref{eq:Tzfit} on the parameters $d$ and $\beta$ for the functions shown in (\textbf{d}) and (\textbf{e}).}
\end{figure*}

\begin{figure}
\centering
\includegraphics[width=\linewidth]{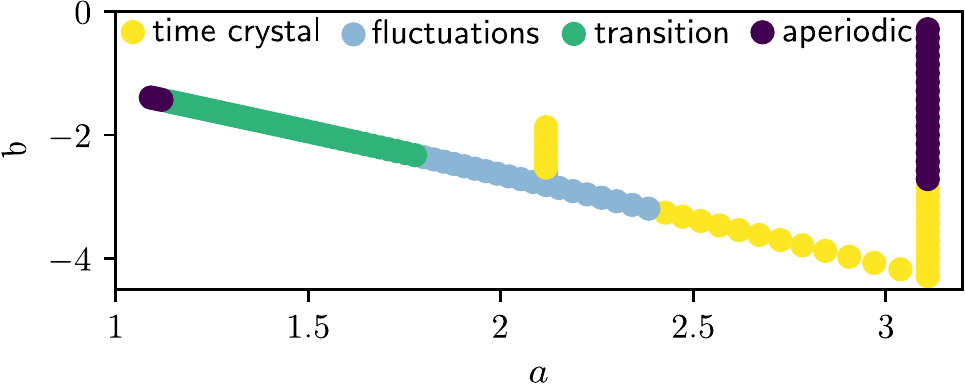}
\caption{\label{fig:interactionsVariation}\textbf{Occurrence of time crystals depending on particle interactions.} The data show the values of the parameters $a$ and $b$ from Eq.\ \eqref{eq:Tzfit} and correspond to the simulations for Fig.~\ref{fig:phaseTransition} and additional simulations that we performed for $d=2.4R$ and various values of $\beta$.}
\end{figure}
In Fig.~\ref{fig:addedInteraction}a, we can see that the swimmers' hydrodynamic interactions mainly influence their nearest neighbors. The torque applied on a second nearest neighbor is, on average, only $1.5\%$ of the one applied on a first nearest neighbor. Moreover, the self-interaction strength of the outer particles is comparable to the interactions with their nearest neighbors.
In the case of the central particle, in contrast, the self-interaction is very small.

Studying the stability of the fixed point that is formed by the motion in the plane allows understanding why the particles can rotate out of the plane for particular parameter configurations. The strength of the torque $T_\mathrm{rot}$ that rotates a particle back into the plane (or further out of the plane) for small deviations from the plane is shown in Figs.\ \ref{fig:addedInteraction}b and \ref{fig:addedInteraction}c. Here, positive values mean a rotation towards the plane whereas negative torques lead to a rotation further out of the plane.
The dependence of this torque on the distance $d$ is shown in Fig.~\ref{fig:addedInteraction}b. Here, we observe large values for small distances. For increasing $d$, the torque decreases and it converges towards $0$. The dependence on $\beta$, which is depicted in Fig.~\ref{fig:addedInteraction}c, shows relatively large stabilizing torques for the time-crystalline state that decrease approximately linearly and become negative for positive values of $\beta$. The values of $T_\mathrm{rot}$ are small compared to the torques $T_z$ that cause a rotation within the plane and not much larger than the numerical uncertainties that lead to the visible fluctuations in the curves. 
These observations show that the periodic motion within the plane is stable for small distances $d$ and small values of $\beta$ but the stability decreases when the parameters increase.

Figures \ref{fig:addedInteraction}d and \ref{fig:addedInteraction}e show our analysis of the pairwise interaction of the inner particle and one of the outer particles in the plane. We see that the distance $d$ between the particles determines the strength of the interactions and that the choice of the propulsion mechanism via the value of $\beta$ mainly determines the shape of the function $T_z(\alpha)$.
From the form of the curves it can be seen that the torque $T_z(\alpha)$, which the inner particle exerts on one of the outer particles, can be described by a function %
\begin{equation}
T_z(\alpha) = a\sin(\alpha)+b\sin(2\alpha) 
\label{eq:Tzfit}%
\end{equation}
with two fit parameters $a$ and $b$. We use nonlinear least squares to obtain the values of the fit parameters for each combination of the values of $d$ and $\beta$ (see Tab.\ \ref{tab:2} in the Appendix). 
For both parameters $a$ and $b$, we find an inverse dependence on $d$ (see Figs.\ \ref{fig:addedInteraction}f and \ref{fig:addedInteraction}g), which can be described by a function 
of the form $1/(c_0 +c_1 d)$. 
The parameter $a$ does not depend on $\beta$, whereas the parameter $b$ turned out to be proportional to $\beta$ (see Fig.\ \ref{fig:addedInteraction}h). 
For the dependence of $a$ and $b$ on $d$ and $\beta$ we thus found the relations
\begin{align}
a(d,\beta)&=\frac{1}{c_{\mathrm{a},0} + c_{\mathrm{a},1} d}\eqcomma\\
b(d,\beta)&=\frac{\beta}{c_{\mathrm{b},0} + c_{\mathrm{b},1} d}\eqdot
\end{align}
By a further nonlinear least squares fit, we obtained the following values for the fit parameters $c_{\mathrm{a},0}$, $c_{\mathrm{a},1}$, $c_{\mathrm{b},0}$, and $c_{\mathrm{b},1}$: 
\begin{align}
c_{\mathrm{a},0} &= -1.335\pm0.004\,\mathrm{Pe}\eta R^2 B_1 \eqcomma \\
c_{\mathrm{a},1} &= 0.753\pm0.002\,\mathrm{Pe}\eta R B_1 \eqcomma \\
c_{\mathrm{b},0} &= -3.350\pm0.006\,\mathrm{Pe}\eta R^2 B_1 \eqcomma \\
c_{\mathrm{b},1} &= 1.841\pm0.003\,\mathrm{Pe}\eta R B_1 \eqdot
\end{align}
Thus, the total torque that the inner particle exerts on one of the outer particles and its dependence on $\alpha$, $d$, and $\beta$ are given by the function 
\begin{equation}
T_z(\alpha,d,\beta) = \frac{1}{c_{\mathrm{a},0} + c_{\mathrm{a},1}d}\sin(\alpha) 
+ \frac{\beta}{c_{\mathrm{b},0} + c_{\mathrm{b},1}d}\sin(2\alpha) \eqdot
\label{eq:Tztotal}%
\end{equation}

To investigate whether time crystals also form for other interactions, we visualize the dependence of the state of the system on the values of the parameters $a$ and $b$ from Eq.\ \eqref{eq:Tzfit} in Fig.\ \ref{fig:interactionsVariation}. Here, we added some further simulations that extended the range of parameters where we found time crystals. Thus, we can conclude that the interactions have a strong influence on the state of the system but time crystals can be found for various different interactions.

\subsection{Larger systems}
Fourth, the time-crystallinity must also hold for larger systems. This criterion distinguishes a time crystal from a simple periodic motion, e.g., of a harmonic oscillator. We confirmed this by increasing the system size. 

\begin{figure}[htb]
\centering\includegraphics[width=\linewidth]{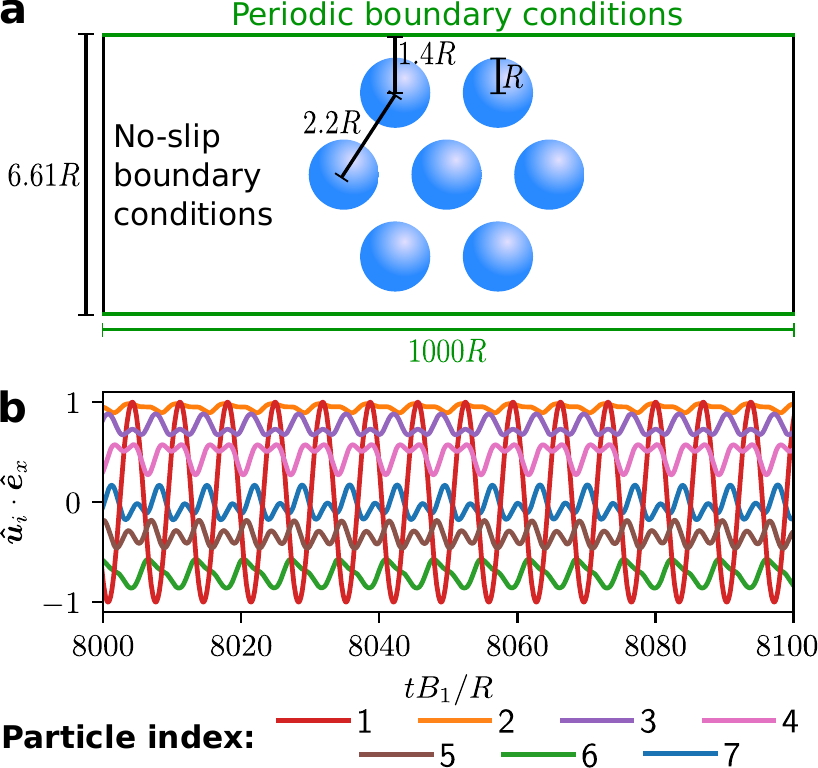}
\caption{\label{fig:geometry}\textbf{Time crystallinity in larger systems.} \textbf{a} Setup for the simulations of an infinite system. \textbf{b} Corresponding simulation results for $d=2.2R$ and $\beta=-3$.}
\end{figure}

Therefore, we extended the particle assembly described in Section \ref{sec:periodicMotion} to an infinite system. 
For this purpose, we reduced the size of the cubic domain in the $y$-direction and introduced periodic boundary conditions. The modified setup is shown in Fig.\ \ref{fig:geometry}a. 
To speed up the calculations and considering that the rotation of the particles in the time crystal shown in Fig.\ \ref{fig:1} takes place completely in the plane, we restricted the swimmers' rotation to the plane. The results are presented in Fig.~\ref{fig:geometry}b. We observed no qualitative changes compared to the simulation results described in Section \ref{sec:periodicMotion}. However, we observed some changes in the way that the outer particles move. These changes are induced by the interactions with the newly added particles that create the larger system. As the system still forms a qualitatively similar motion, this criterion for a time crystal is fulfilled.

These considerations together with the ones in the previous sections show that the time-periodic state of the considered active colloidal crystallite can be identified as a time crystal.

\subsection{Temporal stress-strain diagram}
\begin{figure*}[htb]
\centering\includegraphics[width=\linewidth]{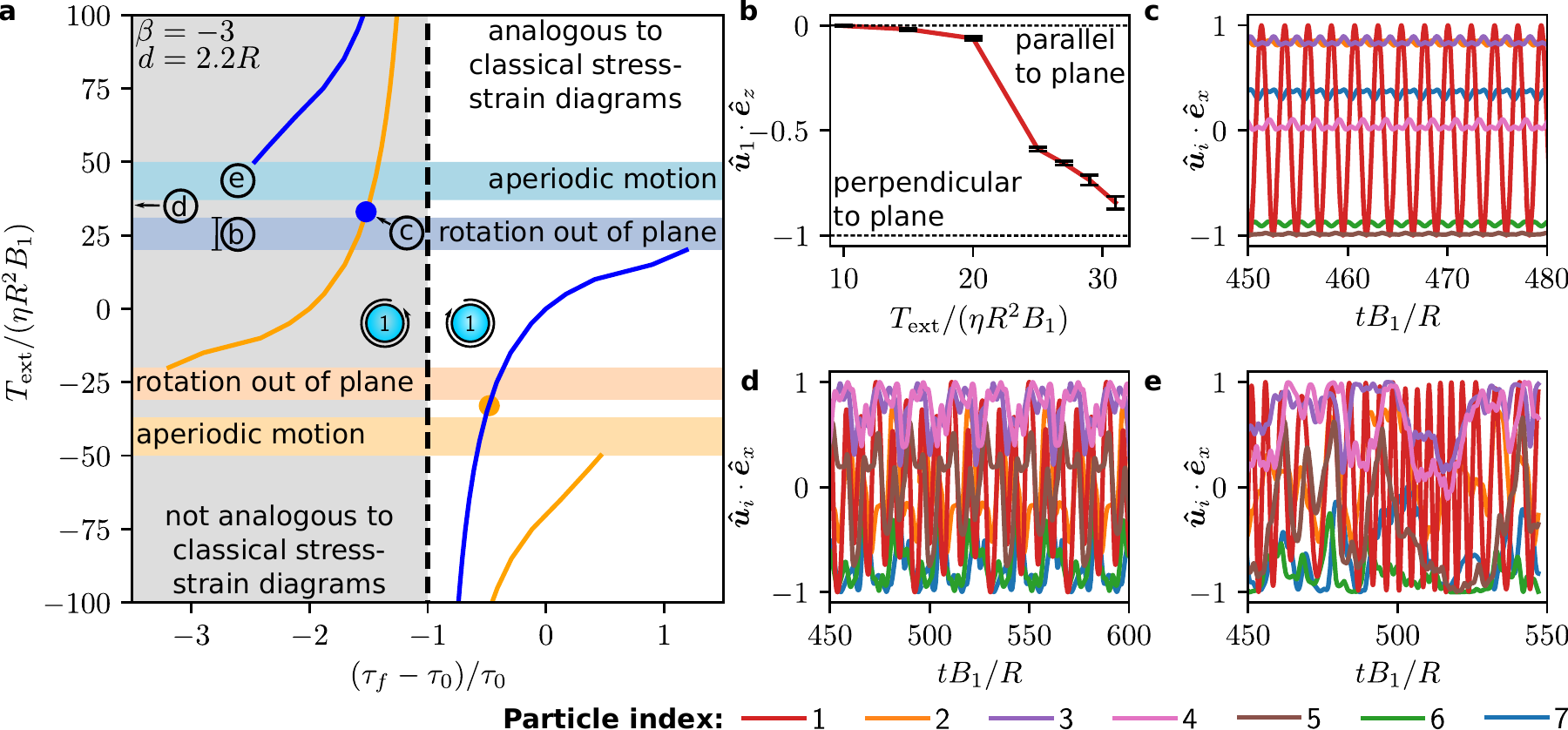}
\caption{\label{fig:4}\textbf{Tempomechanical properties of a colloidal time crystal.} \textbf{a} The temporal stress-strain diagram of the time crystal from Fig.\ \ref{fig:1} describes the period length $\tau_f$ of the oscillation of the time crystal for various external torques $T_{\mathrm{ext}}$ and consists of two parts of which one is in line with the stress-strain diagrams of classical materials and one is not.
\textbf{b} In the interval marked in \textbf{a}, the orientation $\vec{\hat{u}}_1$ of the central particle deviates from the particle plane.
\textbf{c-e} A representative part of the time evolution of the particle orientations $\vec{\hat{u}}_i$ with $i=1,\dotsc,7$ is shown for $T_\mathrm{ext}=33\eta R^2B_1$ (\textbf{c}), $T_\mathrm{ext}=35\eta R^2B_1$ (\textbf{d}), and $T_\mathrm{ext}=37\eta R^2B_1$ (\textbf{e}) with the liquid's viscosity $\eta$.}
\end{figure*}
A time crystal shows strong parallels to classical space crystals, but forms a separate state of matter. The term \ZT{time crystal} was originally proposed as a analogy to breaking the spatial translational symmetry in classical crystals. Thus, it is interesting to see if further analogies besides the translational symmetry breaking can be found. Studying these analogies might create a more descriptive understanding of the time-crystalline behavior and also provides a first step towards finding universal properties that hold for crystallinity in different dimensions.

In classical mechanics, it is common to study the mechanical properties of a material by measuring its stress-strain diagram \cite{moosbrugger2002atlas}. It allows to obtain insights into how the material behaves if an external stress is applied and allows to characterize the material's properties. Transferring this method to active time crystals by introducing a temporal stress-strain diagram would allow us to find parallels and differences to classical materials and to deduce information about the material properties of this novel state of matter.

Therefore, we here investigate the tempomechanical properties of the observed colloidal time crystal by determining its temporal stress-strain diagram. For this purpose, we apply external torques $T_{\mathrm{ext}}$ to the central particle and determine in each case how the period length $\tau_f$ of the oscillation changes relative to the period length $\tau_0$ for no external torque.
Like the torque exerted on the central particle by its interaction with the other particles, the external torque is perpendicular to the particle plane (both torques are parallel or antiparallel). Based on the analogy to classical crystals and stress-strain diagrams, we can consider the torque $T_{\mathrm{ext}}$ analogously to the stress in a classical stress-strain diagram. The change in period length $\tau_f-\tau_0$ is the answer of the system and corresponds to the strain in the classical case.
When the central particle rotates in the opposite direction compared to the case without external torque, we add a negative sign to $\tau_f$. It should be noted that a negative change in the period length has no classical analog.

The behavior of the system considered in this work is invariant regarding the rotation of the whole system by an angle of $180^\circ$ about either of the coordinate axes. Introducing an external torque, however, breaks this symmetry. Now, in case of a rotation about the $x$- or $y$-axis by $180^\circ$, the rotation direction of the particles is reversed. Thus, there are two equivalent possible setups when $T_{\mathrm{ext}}$ is nonzero. These two setups lead to two curves describing the relation of $T_{\mathrm{ext}}$ and $(\tau_f-\tau_0)/\tau_0$. The two curves are equivalent and can be mapped onto each other by a point reflection at $(T_{\mathrm{ext}},\tau_f)=(0,0)$. 
In the following, we study the tempomechanical properties of the time crystal observed in Fig.\ \ref{fig:2} based on a temporal stress-strain diagram (see Fig.\ \ref{fig:4}). Thereby, we focus on the curve with a clockwise rotation of the central particle for $T_{\mathrm{ext}}=0$ (the blue curve in Fig.\ \ref{fig:4}a), since the observations for the other curve are completely analogous.

The temporal stress-strain curve corresponding to the time-crystalline state shows a nonlinear-elastic behavior, which also occurs in many classical materials like plastic. 
For negative and decreasing $T_{\mathrm{ext}}$, $\tau_f$ declines and asymptotically approaches $\tau_f=0$, which corresponds to the mechanic behavior of an ordinary material under compression. Also, for large and increasing $T_{\mathrm{ext}}$, the curve asymptotically approaches $\tau_f=0$, but now it approaches from negative values of $\tau_f$, which has no analogy to the mechanical behavior of ordinary materials. 
The asymptotic behavior for very large and very small values of $T_{\mathrm{ext}}$ results from the fact that the external torque dominates the torques exerted by the other particles on the central particle. Thus, it determines the rotation of the central particle for large $\abs{T_{\mathrm{ext}}}$. 
When $T_{\mathrm{ext}}$ is small and positive, $\tau_f$ increases with $T_{\mathrm{ext}}$, which corresponds to the mechanic behavior of an ordinary material under stretching. 
For approaching intermediate positive values of $T_{\mathrm{ext}}$ from above, the curve runs towards strongly negative values of $\tau_f$, whereas when approaching the intermediate values from below, the curve runs towards large positive values of $\tau_f$. This behavior is in line with the fact that at some intermediate external torque, the external torque cancels the torque exerted on the central particle by interactions with the other particles. Here, one would expect a divergence of $\tau_f$. 
The cancellation can be estimated to occur at $T_{\mathrm{ext}}\approx 35\eta R^2B_1$, since the torque exerted by the other particles on the central one for $T_{\mathrm{ext}}=0$ is about $-35\eta R^2B_1$. 

However, in the range between $T_{\mathrm{ext}}=20\eta R^2B_1$ and $T_{\mathrm{ext}}=50\eta R^2B_1$, the behavior of the time crystal is in fact much more complicated. In this range, the time-crystalline behavior changes qualitatively or vanishes. 
For values between $T_{\mathrm{ext}}=20\eta R^2B_1$ and $T_{\mathrm{ext}}=31\eta R^2B_1$, we saw that the orientation of the central rotating particle is no longer within the particle plane. Its deviation from the plane increases with $T_{\mathrm{ext}}$ (see Fig.\ \ref{fig:4}b). 
We also observed periodic high-frequency fluctuations disturbing the angle between the orientation of the central particle and the plane. 
The standard deviation of this disturbance is shown in Fig.\ \ref{fig:4}b as error bars. 
One can consider the behavior in this interval of $T_{\mathrm{ext}}$ as a different state of the time crystal. 
Applying a larger external torque $T_{\mathrm{ext}}=33\eta R^2B_1$ leads to a completely different behavior. After a relatively long aperiodic transition phase, the particles rotate again periodically within the particle plane (see Fig.\ \ref{fig:4}c). 
This state occurs at a point in the temporal stress-strain diagram that corresponds to a point lying on the second (orange) stress-strain curve. Thus, applying this torque destabilizes the motion enough to flip the system to its other equivalent setup. Indeed, the central particle now rotates in the opposite direction than for lower values of $T_{\mathrm{ext}}$. 
The total torque acting on this particle is then $67.5\eta R^2B_1$, which equals approximately the externally applied torque $T_{\mathrm{ext}}=33\eta R^2B_1$ plus the torque $35\eta R^2B_1$ exerted on the central particle by the other ones when the system is not influenced externally. Thus, the point lies on the curve created by a point reflection of the blue curve at $(T_{\mathrm{ext}},\tau_f)=(0,0)$. From this reasoning, we can deduce that the behavior described by the blue curve looses its stability in this point. 
Near $T_{\mathrm{ext}}=35\eta R^2B_1$, the total torque acting on the central particle is approximately zero. Then, the motion is still periodic, but it has a rather large period length $\tau_f$ and the motion within a period is rather complex (see Fig.\ \ref{fig:4}d and supplemental video V5). 
Finally, in the region between $T_{\mathrm{ext}}=37\eta R^2B_1$ and $T_{\mathrm{ext}}=50\eta R^2B_1$, the system shows an aperiodic motion that can be interpreted as a fractured time crystal (see Fig.\ \ref{fig:4}e). This state can be seen as an analogy to a broken crystal. When $T_{\mathrm{ext}}$ increases, the aperiodicity decreases until the system shows a periodic motion again for $T_{\mathrm{ext}}=50\eta R^2B_1$ and larger external torques. However, in this case, the central particle rotates in the opposite direction as for large negative external torques. In this region, the external torque starts dominating the system's behavior again. Here, a comparison to classical crystals is not reasonable. While a fractured time crystal can find its way back to a periodic state, a fractured classical crystal will not become periodic again when a larger stress is applied. Though, it should be noted that this state is not a truly time-periodic state as the periodicity is induced by the external torque.

\section{Discussion}
We observed a continuous time-crystalline behavior of an active colloidal crystallite and analyzed the occurrence and tempo-mechanical properties of this exceptional state of matter. Such colloidal time crystals form a fascinating new class of materials that deserves extensive exploration in future research. 

Our investigation revealed requirements for the occurrence of a colloidal time crystal. 
It showed that an assembly of squirmers as studied in this work self-assembles a classical continuous time crystal when the values of the particle distance and squirming parameter are sufficiently small. We found that the occurrence of this state is neither prevalent nor limited to a selection of very particular parameter values, that it is robust against fluctuations originating from numerical errors or Brownian motion, that it strongly depends on the interactions of the particles, and that it occurs also for larger systems.
Considering different particle assemblies, we found that the spatial assembly of the particles has a strong influence on their hydrodynamic interactions and thus on their potential formation of a time-crystalline state.

In the future, this work should be continued by additional simulation studies that consider the dependence of the behavior of the system on the parameter values in greater detail, calculate the state diagram with higher resolution and for larger ranges of the parameter values, and take also qualitatively different particle assemblies into account. In the latter case, it would be interesting to study various regular and irregular spatial assemblies of active particles in two and three spatial dimensions.
Our analytic representation of the hydrodynamic interactions of two particles provides an ansatz for replacing the computationally expensive hydrodynamic simulations performed in this work by much more efficient Brownian dynamics simulations that could strongly speed up future research. 
With Eq.\ \eqref{eq:Tztotal} that we provided for the interaction torque, it should be possible to perform Brownian dynamics simulations for an infinitely large hexagonal crystal of active particles. 
For other particle assemblies, including those studied in the present work, a similar approach should be possible, but in this case it would be necessary to first determine also fit functions for the interaction torques exerted by a particle at the boundary of the crystallite on other particles. 

Another important future step in the investigation of colloidal time crystals is their experimental realization. For this, our numerical work lays a good basis by providing guidelines how a system of active particles must be designed to be able to exhibit time-crystalline behavior. 
The realization of the system studied in the present article is possible by arranging and holding biological or artificial microswimmers on a hexagonal lattice with optical or acoustical tweezers \cite{TakatoridRVB2015}. 
Particles that generate suitable flow fields are those that are well described by the pusher model. This is the case for microorganisms like \textit{Escherichia coli} or artificial microswimmers like Janus particles \cite{brown2016swimming}. Recently, it was also shown that particles driven by ultrasound create a flow field similar to the one described by the pusher model \cite{voss2020shape}. These particles allow an easy tracking of the particle directions, e.g., by observing the system in a microscope \cite{buttinoni2013dynamical, volpe2011microswimmers, das2015boundaries}. 
Regarding the assembly of the particles, we propose to use acoustic tweezers, since they are not affected by the inhomogeneous optical properties that are common for active particles. 
Light fields should preferably be used to supply light-propelled active particles by energy \cite{SafdarSJ2017}. 
The spatial configuration of the microswimmers that we studied in our work could also be achieved by self-assembly when the particles are equipped with suitable interactions. 
An option to equip artificial active particles with repulsive interactions that can cause a hexagonal particle assembly is to embed a magnetic core within the particles \cite{ebert2009experimental}. By varying the strength of an external magnetic field, which is oriented perpendicular to the particle plane and induces a magnetic dipole in the particles, it is then possible to tune the repulsive interactions of the particles and thus the lattice constant.

\section{Methods}
\subsection{Model for microswimmers}
We model the active particles as spherical squirmers as proposed by Lighthill \cite{lighthill1952squirming} and Blake \cite{blake1971spherical}. This model is widely used for various microswimmers ranging from biological ones like \textit{Paramecia} to artificial ones like Janus particles.
The distinction between pullers, neutral squirmers, and pushers is possible via the squirming parameter $\beta = B_1/B_2$, where $B_1$ and $B_2$ are the first and second velocity mode, respectively. In this work, we choose the values $\beta=-3,0,3$ for the squirming parameter to be consistent with Ref.\ \onlinecite{theers2016modeling}. To facilitate the treatment of the spherical particles, spherical coordinates $r,\theta,\phi$ are used. The particle surface is located at $r=R$, where $R$ is the radius of the particles. On the surface of the particles, the velocity of the surrounding liquid is prescribed as
\begin{equation}
\boldsymbol{v}_{\text{sq}}(R, \theta, \varphi) = B_1(1+\beta\cos(\theta))\sin(\theta)\boldsymbol{\hat{s}}\eqdot
\label{eq:boundary1}%
\end{equation}
Here, $\boldsymbol{\hat{s}}$ is the tangential vector on the surface and equivalent to the unit vector $\boldsymbol{\hat{e}}_{\theta}$.
The analytical solution of the hydrodynamic problem leads to a swimming velocity 
\begin{equation}
U_0=\frac{2}{3}B_1\eqdot
\label{eq:swimmingVelocity}%
\end{equation}

\subsection{Hydrodynamics}
The active particles generate a flow field in the surrounding liquid that leads to hydrodynamic interactions of the particles, which affect their motion.  
Since the size of microswimmers typically lies in the range of micrometers, the Reynolds number $\text{Re}=\rho R U_0/\eta$ with the liquid's mass density $\rho$, maximal flow speed $U_0$, and dynamic shear viscosity $\eta$ is much smaller than one. 
This allows to approximate the Navier-Stokes equation describing the flow field by the Stokes equation \cite{dhont1996introduction}
\begin{equation}
-\eta \Delta \boldsymbol{v}+\Nabla p = \vec{0} \eqcomma 
\label{eq:stokes1}%
\end{equation}
where $\boldsymbol{v}$ denotes the liquid's velocity field and $p$ its pressure field. 
Any external forces applied to the system, such as gravity, can be neglected in this work. 
Additionally, the fluid is treated as incompressible:
\begin{equation}
\Nabla \cdot \boldsymbol{v} = 0\eqdot
\label{eq:stokes2}%
\end{equation}
From the solution of the Stokes equation \eqref{eq:stokes1}, one can calculate the stress tensor \cite{BechingerdLLRVV2016}
\begin{equation}
\boldsymbol{\Sigma}=-p\boldsymbol{I}+\eta\left(\Nabla\otimes\boldsymbol{v}+(\Nabla\otimes\boldsymbol{v})^{\mathrm{T}}\right)\eqcomma
\label{eq:stress}%
\end{equation}
where $\boldsymbol{I}$ is the identity matrix, $\otimes$ is the dyadic product, and $^\mathrm{T}$ denotes transposition. 

Since the positions of the particles are fixed, they can only rotate but not translate. Therefore, we can neglect the forces and need to consider only the torques exerted on the particles by hydrodynamic interactions, when studying their motion. 
Based on the stress tensor \eqref{eq:stress}, the hydrodynamic torque on a particle can be calculated as \cite{BechingerdLLRVV2016}
\begin{equation}
\boldsymbol{T^h}=\int_S\boldsymbol{r}\times(\boldsymbol{\Sigma}\cdot\boldsymbol{\hat{n}})\,\dif s\eqcomma
\label{eq:torque}%
\end{equation}
where $\boldsymbol{r}$ is the position of a surface element with area $\dif s$ and normal vector $\boldsymbol{\hat{n}}$ on the particle surface $S$.

\subsection{Equation of motion}
Starting from Newton's equation of motion, one obtains for the considered system the Langevin equation
\begin{equation}
\boldsymbol{\dot{J}}=-\gamma_R\boldsymbol{\omega}+\boldsymbol{T^h}+\gamma_R\boldsymbol{\xi}_R  
\end{equation}
that describes the motion of an active particle. 
Here, $\boldsymbol{J}$ is the angular momentum of the particle, $\gamma_R$ its rotational friction coefficient, $\boldsymbol{\omega}$ its angular velocity, and $\boldsymbol{\xi}_R(t)$ is a unit-variance Gaussian white noise random torque that takes Brownian rotation into account.  
$\gamma_R$ is related to the rotational diffusion coefficient of the particle by $D_R=k_\mathrm{B} T/\gamma_R$, where $k_\mathrm{B}$ is the Boltzmann constant and $T$ the temperature.

In the regime of small Reynolds numbers, the rotational friction is large compared to $\boldsymbol{\dot{J}}$. Thus, inertia can be neglected leading to
\begin{equation}
\vec{0}=-\gamma_R\boldsymbol{\omega}+\boldsymbol{T^h}+\gamma_R\boldsymbol{\xi}_R \eqdot
\end{equation} 
This equation represents an energy balance between the energy lost by friction, the energy obtained by the hydrodynamic torque that acts on the particle, and the torque contribution from the Brownian motion.
Using additionally the relation 
\begin{equation}
\boldsymbol{\omega}=\boldsymbol{\hat{u}}\times \frac{\dif\boldsymbol{\hat{u}}}{\dif t} 
\end{equation}
with the particle director $\boldsymbol{\hat{u}}$ (a unit vector that describes the particle's orientation), one obtains
\begin{equation}
\boldsymbol{\dot{\hat{u}}}= \bigg(  \frac{1}{\gamma_R}\boldsymbol{T^h} +\boldsymbol{\xi}_R\bigg) \times\boldsymbol{\hat{u}}\eqdot
\label{eq:langevinRotation}%
\end{equation}

If the swimming velocity is high and, therefore, the dimensionless P\'eclet number $\text{Pe}=B_1/(R D_R)$ is large, the Brownian motion is also negligible and we get
\begin{equation}
\boldsymbol{\dot{\hat{u}}}=\frac{1}{\gamma_R}\boldsymbol{T^h}\times\boldsymbol{\hat{u}}\eqdot
\end{equation}

In the simulations where an external torque $\boldsymbol{T}_{\mathrm{ext}}$ is taken into account, we simply add $\boldsymbol{T}_{\mathrm{ext}}=T_{\mathrm{ext}}\boldsymbol{\hat{e}}_{z}$ to $\boldsymbol{T^h}$ in the equations above, where $T_{\mathrm{ext}}$ is the scalar torque and $\boldsymbol{\hat{e}}_{z}$ the unit vector corresponding to the $z$ axis.

\subsection{\label{sec:Implementation}Implementation}
For the numerical solution of Eqs.\ \eqref{eq:stokes1} and \eqref{eq:stokes2}, the \texttt{FEniCS} computing platform \cite{kirby2004algorithm,kirby2006compiler,logg2010dolfin, olgaard2010optimizations,alnaes2014unified,alnaes2015fenics,logg2012automated}
applying the finite element method is used. To implement the equations in their weak formulation 
\begin{gather}
\begin{split}
&\int_{\Omega}\eta (\Nabla\otimes\boldsymbol{f}_t):(\Nabla\otimes\boldsymbol{v})\,\dif ^{3}x 
+\int_{\Omega}p\Nabla\cdot\boldsymbol{f}_t\,\dif^{3}x  \label{eq:stokesWeakI} \\
&= \int_{\partial \Omega} \!\!\! \boldsymbol{f}_t\cdot\boldsymbol{t}\,\dif s  \eqcomma
\end{split}\\
\int_{\Omega}q_t\Nabla \cdot \boldsymbol{v}\,\dif^{3}x = 0\eqcomma
\label{eq:stokesWeakII}%
\end{gather}
the Python interface is applied. 
Here, $\Omega$ describes the simulation domain and $\partial\Omega$ its surface, $\boldsymbol{f}_t$ and $q_t$ are test functions, and 
$\boldsymbol{t}=\eta(\Nabla\otimes\boldsymbol{v})\cdot\boldsymbol{\hat{n}}+p\boldsymbol{\hat{n}}$ is a so-called pseudotraction \cite{donea2003finite}.

In our main simulations, a liquid-filled cube-shaped domain with an edge length of $1000R$ and no-slip boundary conditions is used. The particles are located in the center of the cubic domain with a distance of $d$ to their nearest neighbors. We vary the distance between the particles between $d=2.2R$ and $d=3R$. The distance between the particles and the boundaries of the cubic domain is large enough to make the influence of the boundary conditions on the particle motion negligible. For the boundary conditions on the particle surfaces, the squirmer model \eqref{eq:boundary1} is used.

As a solver for the linear systems of equations involved in numerically treating Eqs.\ \eqref{eq:stokesWeakI} and \eqref{eq:stokesWeakII}, the minimal residual method \texttt{minres} is chosen. Additionally, the algebraic multigrid preconditioner \texttt{amg} is used. To speed up the calculations, the preconditioner $\int_\Omega((\Nabla\otimes\vec{u}):(\Nabla\otimes\vec{v}_t)+pq_t) \,\dif ^{3}x$ is applied.

We create a tetrahedral mesh with the program \texttt{gmsh} \cite{geuzaine2009gmsh}. To shorten computation times, an adaptive mesh is generated. The maximal distance between nodes is $0.1R$ on the particle surface and $50R$ on the boundary of the cubic domain. There is a refined region around the particles with a maximal node distance of $0.4R$. This region ends at a distance of $2R$ to all particles.
The total number of tetrahedra in the mesh is between $141530$ for $d=2.2R$ and $161006$ for $d=3R$.

As an initial condition, the orientations of the particles are set randomly using the function \texttt{numpy.random.rand} with a random seed of $0$. To ensure that the observed results do not depend on the initial orientations, we simulated the system for $d=2.2R$ in total $10$ times with random seeds from $0$ to $9$ and compared the results.

\begin{figure}[htb]
\centering\includegraphics[width=\linewidth]{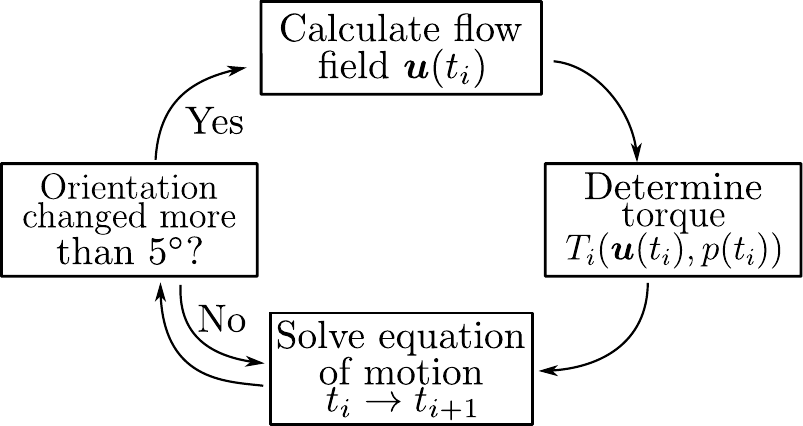}
\caption{\label{fig:algorithm}\textbf{Flow diagram of the algorithm used for simulating active crystallites.}}
\end{figure}

The equation of motion \eqref{eq:langevinRotation} is solved for each swimmer individually by using the Euler-Maruyama method~\cite{maruyama1955continuous}.
Thus, the time-evolution of the particle orientations is calculated by the scheme
\begin{equation}
\begin{split}%
\boldsymbol{\hat{u}}(t+\Delta t) 
&=\boldsymbol{\hat{u}}(t)+\Delta t\left(\frac{D_R}{k_\mathrm{B} T}\boldsymbol{T^h}(t)\times\boldsymbol{\hat{u}}(t)\right)\\
&\quad\:\! +\sqrt{\Delta t}\left(\boldsymbol{\xi}_R(t)\times\boldsymbol{\hat{u}}(t)\right)\eqcomma
\end{split}%
\end{equation}
where the time step size $\Delta t=10^{-2}B_1/R$ is used. 
Since the calculations of the flow field are computationally expensive, we reduce their number by using an iterative simulation approach represented schematically in Fig.\ \ref{fig:algorithm}.
After each Euler step, it is checked if at least one particle moved more than \SI{5}{\degree} since the last calculation of the flow field. In that case, the flow field and the torques applied to the particles are recalculated as described above. If the particles moved less, the next Euler step is determined directly using the previously calculated torque as an approximation. For the simulations used in Figs.\ \ref{fig:phaseTransition} and \ref{fig:addedInteraction}, we decreased the threshold for recalculations to $2^\circ$ to obtain smoother curves for the analysis of interactions. 
The simulations are run for a duration of at least $600B_1/\sigma$.

For calculating a stress-strain diagram as well as the results presented in Figs.\ \ref{fig:phaseTransition} and \ref{fig:addedInteraction}, the initial particle orientations are chosen as one of the orientations occurring in the periodic motion for no external torque. The corresponding particle directors are given in Tab.\ \ref{tab:particleDirectors}.
\begin{table}[htb]
\centering
\caption{\label{tab:particleDirectors}\textbf{Initial configuration corresponding to the stress-strain diagram in Fig.\ \ref{fig:4}.}}
\begin{tabular}{|c|c|}
\hline
\textbf{Particle index }$\boldsymbol{i}$ & \textbf{Particle director }$\vec{\hat{u}_i}$ \\ \hline
$1$ & $(-0.22, -0.97, 0.00)^{\mathrm{T}}$ \\
$2$ & $(0.10, 0.99, 0.00)^{\mathrm{T}}$ \\
$3$ & $(0.62, 0.78, 0.00)^{\mathrm{T}}$ \\
$4$ & $(0.87, -0.48, 0.00)^{\mathrm{T}}$ \\
$5$ & $(0.35, -0.93, 0.00)^{\mathrm{T}}$ \\
$6$ & $(-0.84, -0.54, 0.00)^{\mathrm{T}}$ \\
$7$ & $(-0.97, 0.32, 0.00)^{\mathrm{T}}$ \\
\hline
\end{tabular}
\end{table}

\begin{widetext}
\begin{acknowledgments}
We thank Michael te Vrugt for helpful discussions. 
R.W.\ is funded by the Deutsche Forschungsgemeinschaft (DFG, German Research Foundation) -- WI 4170/3-1. 
The simulations for this work were performed on the computer cluster PALMA II of the University of M\"unster. 
\end{acknowledgments}

\appendix
\section{Fit-parameter values for Eq.\ (\ref{eq:Tzfit})}
Our results for the fit parameters $a$ and $b$ that we obtained by fitting Eq.\ \eqref{eq:Tzfit} to our simulation data for the interaction torque $T_z$ are shown in Tab.\ \ref{tab:2}.
\begin{table*}[htbp]
\begin{center} 
\caption{\label{tab:2}\textbf{Fit-parameter values for the interaction torque.} The values of the fit parameters $a$ and $b$ from Eq.\ \eqref{eq:Tzfit} as well as their standard deviations $\Delta a$ and $\Delta b$ are given for various values of the parameters $d$ and $\beta$.}
\begin{tabular}{|c|c|c|c|c|c|} 
\hline
$d/R$ & $\beta$ & $a/(\mathrm{Pe}\eta R^2 B_1)$ & $\Delta a/(\mathrm{Pe}\eta R^2 B_1)$ & $b/(\mathrm{Pe}\eta R^2 B_1)$ & $\Delta b/(\mathrm{Pe}\eta R^2 B_1)$ \\ 
\hline
$2.20$ & $ -0.5$ & $3.1197$ & $0.0005$ & $-0.7116$ & $0.0005$ \\ 
$2.20$ & $ -1.0$ & $3.1225$ & $0.0005$ & $-1.4285$ & $0.0005$ \\ 
$2.20$ & $ -1.5$ & $3.1202$ & $0.0008$ & $-2.1422$ & $0.0008$ \\ 
$2.20$ & $ -1.6$ & $3.1236$ & $0.0006$ & $-2.2845$ & $0.0006$ \\ 
$2.20$ & $ -1.7$ & $3.1232$ & $0.0007$ & $-2.4278$ & $0.0007$ \\ 
$2.20$ & $ -1.8$ & $3.1218$ & $0.0006$ & $-2.5694$ & $0.0006$ \\ 
$2.20$ & $ -1.9$ & $3.1158$ & $0.0005$ & $-2.7089$ & $0.0005$ \\ 
$2.20$ & $ -2.0$ & $3.1187$ & $0.0007$ & $-2.8567$ & $0.0007$ \\ 
$2.20$ & $ -2.1$ & $3.1219$ & $0.0005$ & $-2.9973$ & $0.0005$ \\ 
$2.20$ & $ -2.2$ & $3.1192$ & $0.0004$ & $-3.1369$ & $0.0004$ \\ 
$2.20$ & $ -2.3$ & $3.1277$ & $0.0013$ & $-3.2893$ & $0.0013$ \\ 
$2.20$ & $ -2.4$ & $3.1246$ & $0.0008$ & $-3.4246$ & $0.0008$ \\ 
$2.20$ & $ -2.5$ & $3.1205$ & $0.0011$ & $-3.5649$ & $0.0011$ \\ 
$2.20$ & $ -3.0$ & $3.1244$ & $0.0009$ & $-4.2829$ & $0.0009$ \\ 
$2.25$ & $ -3.0$ & $2.7760$ & $0.0007$ & $-3.7676$ & $0.0007$ \\ 
$2.30$ & $ -3.0$ & $2.5176$ & $0.0007$ & $-3.3822$ & $0.0007$ \\ 
$2.35$ & $ -3.0$ & $2.2949$ & $0.0006$ & $-3.0591$ & $0.0006$ \\ 
$2.40$ & $ -3.0$ & $2.1120$ & $0.0005$ & $-2.8018$ & $0.0005$ \\ 
$2.45$ & $ -3.0$ & $1.9570$ & $0.0004$ & $-2.5804$ & $0.0004$ \\ 
$2.50$ & $ -3.0$ & $1.8262$ & $0.0003$ & $-2.3968$ & $0.0003$ \\ 
$2.55$ & $ -3.0$ & $1.7125$ & $0.0005$ & $-2.2375$ & $0.0005$ \\ 
$2.56$ & $ -3.0$ & $1.6919$ & $0.0005$ & $-2.2044$ & $0.0005$ \\ 
$2.57$ & $ -3.0$ & $1.6733$ & $0.0003$ & $-2.1748$ & $0.0003$ \\ 
$2.58$ & $ -3.0$ & $1.6503$ & $0.0004$ & $-2.1461$ & $0.0004$ \\ 
$2.59$ & $ -3.0$ & $1.6306$ & $0.0004$ & $-2.1153$ & $0.0004$ \\ 
$2.60$ & $ -3.0$ & $1.6136$ & $0.0003$ & $-2.0928$ & $0.0003$ \\ 
$2.65$ & $ -3.0$ & $1.5224$ & $0.0004$ & $-1.9628$ & $0.0004$ \\ 
$2.70$ & $ -3.0$ & $1.4407$ & $0.0003$ & $-1.8491$ & $0.0003$ \\ 
$2.75$ & $ -3.0$ & $1.3691$ & $0.0003$ & $-1.7517$ & $0.0003$ \\ 
$2.80$ & $ -3.0$ & $1.3047$ & $0.0003$ & $-1.6580$ & $0.0003$ \\ 
$2.20$ & $ 0.0$ & $3.1221$ & $0.0004$ & $-0.0013$ & $0.0004$ \\ 
$2.20$ & $ 3.0$ & $3.1157$ & $0.0012$ & $4.2785$ & $0.0012$ \\ 
\hline 
\end{tabular} 
\end{center}
\end{table*}
\clearpage
\end{widetext}

\bibliographystyle{naturemag}
\bibliography{ref}

\end{document}